\documentclass[pra,twocolumn,aps,amssymb,footinbib,showpacs]{revtex4-1}
\usepackage{amssymb}
\usepackage{amsmath}
\usepackage{times}
\usepackage{latexsym}
\usepackage{graphicx}
\usepackage{color}
\usepackage{bm}
\usepackage{subfigure}
\begin{document}
\title{Topology Induced  Oscillations in Majorana Fermions in a Quasiperiodic Superconducting Chain}
\author{  Indubala I Satija }
\affiliation{Department of Physics, George Mason University , Fairfax, VA 22030, USA}
\date{\today}
\begin{abstract}
{Spatial profile of the Majorana fermion wave function  in a one-dimensional $p$-wave superconductors ($\cal{PWS}$) with quasi periodic disorder is shown to exhibit  spatial oscillations. These oscillations
  damp out  in the interior of the chain and are characterized by a
period that has topological origin  and is equal to the Chern number determining the Hall conductivity near half-filling of  a two-dimensional electron gas in a crystal.  This mapping unfolds in view of a
 correspondence between the critical point for the
topological transition in $\cal{PWS}$ and the {\it strong coupling fixed point} of the Harper's equation.  Oscillatory character of these modes persist in a generalized model 
related to an extended Harper  system where the electrons also tunnel to the diagonals  of a  square lattice.  However, beyond a bicritical point,
the Majorana oscillations  occur with a random period, 
characterized by  an invariant fractal set. }

\end{abstract}

\maketitle

\section{Introduction}

Majorana Fermions -- predicted in  $1937$ by  Ettore Majorana\cite{Maj, SA}, are  charge-neutral fermions that are their own antiparticles. These mysterious particles remain elusive, although there are many speculative  theories about their incarnation  in a variety  of problems that include  neutrino oscillations, supersymmetry, dark matter and also in topological excitations in superconductors\cite{Will}.
Following a pioneering work by Kitaev \cite{Kitaev},
the one-dimensional $p$-wave superconducting quantum wire has emerged as one of the key system for studying
Majorana Fermions.   Kitaev's proposal has broadened the scope of exploration of these particles, beyond the perimeter of  particle physics to the condensed matter, energizing  both
 the study of these exotic particles and also the  field of superconductivity. 
The iconic system underlying these studies
 is given by the following one dimensional Hamiltonian,

\begin{equation}
H_{sc}=\sum_{ n = -\infty}^{\infty}  J (c_{n+1}^{\dagger}c_n^{} + \Delta c_{n+1}^{\dagger} c^{\dagger}_n) + \textrm{c.c.} + \mu (c_n^{\dagger} c_n^{} -1/2).
\label{SC}
\end{equation}

Here $c_n^{\dagger}$ is the fermion creation operator at the site $n$ of the superconducting chain. The parameter $J$ is the nearest-neighbor hopping amplitude for the fermions, $\Delta$ the superconducting gap function (assumed real), and $\mu$ the on-site chemical potential.  Modeling a $\cal{PWS}$, the system exhibits
a quantum phase transition as one tunes $\mu$ from $\mu > J$ (topologically trivial) to $\mu < J$ (topological nontrivial) with $\mu=J$ being the point of  topological transition \cite{Kitaev}.

The above Hamiltonian also describes a spin chain, the transverse field Ising-like model\cite{LM},  

\begin{eqnarray}
H_{spin} =\sum_{ n = -\infty}^{\infty} [ J_x  \sigma^x_n \sigma^x_{n+1}
  + J_y \sigma^y_{n} \sigma^y_{n+1}
  +  \mu  \,\  \sigma^z_n ]
  \label{Ising}
\end{eqnarray}

Here $J_x$ and $J_y$  are the exchange interactions along the $x$ and the $y$ directions in spin space  with spin space anisotropy $J_y-J_x$ and $\mu$ is the transverse magnetic field.
The two models given by Eqs (\ref{SC}) and (\ref{Ising}) are equivalent in view of  the Jordan-Wigner transformation\cite{LM} that maps the spin operators to fermion operators, with $J_x = J(1-\Delta)$ and $J_y= J(1+\Delta)$.  Thus the superconducting gap parameter $\Delta$ in $\cal{PWS}$ plays the role of the spin space anisotropy in the Ising chain.

There has been numerous studies of the spin chain\cite{DS,C} with random or quasi periodic disorder modeled by either the exchange interaction $J$ or the magnetic field $\mu$  being site dependent.
These two ways of introducing inhomogenity in the system are somewhat equivalent reflecting the well known duality of the Ising model model\cite{Fradkin}. 
The eigenvalue equation for the system where $\mu$ is site dependent and denoted  below as $\mu_n$ is given by  the following  set of coupled  equations,

\begin{eqnarray}
(1-\Delta) \psi^a_{n+1}+(1+\Delta)\psi^a_{n-1} +2  \mu_n \psi^a_n  = E \psi^b_n
\label{sup1}\\
(1+\Delta) \psi^b_{n+1}+(1-\Delta) \psi^b_{n-1} + 2  \mu_n \psi^b_n  =  E \psi^a_n
\label{superc}
\end{eqnarray}

Here we have set $J$ equal to unity, that is  $\mu$ and $E$ are scaled by $J$. 

In this paper, we revisit the study of quasi periodic  superconducting chain with sinusoidal modulation in $\mu_n$ described by,
\begin{equation}
\mu_n = 2 \lambda \cos ( 2\pi \phi n + \delta).
\label{qumu}
\end{equation}
Here $\phi$ is an irrational number  and $\delta$ is a phase factor. 
Analogous to a periodic chain, such a quasi periodic system has been shown to exhibit the Ising transition to magnetic long range order\cite{IISD, SN},. For the corresponding superconducting chain,
this corresponds to the topological transition where the  critical value $\lambda_T$  is given by,

\begin{equation}
\lambda_T = 1+\Delta
\end{equation}

In other words, the superconducting chain with quasiperiodic sinusoidal modulation  supports Majorana modes for $ \lambda < \lambda_T = 1+\Delta$.
 
Motivation to study such a quasi periodic system was partially due to the fact that for $\Delta=0$, the  Eq. (\ref{superc}) reduces to a well known tight binding equation  for $\psi=\psi^a=\psi_b$,  known as the Harper equation\cite{Harper}  
 \begin{equation}
\psi_{n+1}+\psi_{n-1} + 2 \lambda \cos ( 2 \pi \phi n+ \delta) \psi_n = E \psi_n,
\label{Harper}
\end{equation}

Harper equation models the iconic problem of a two-dimensional electron gas (2DEG) in a crystal in a magnetic field.  The parameter $\phi$ is the magnetic flux ( measured in units of
flux quantum $\hbar/e$ ).  The system exhibits all possible integer quantum Hall states\cite{QHE, TKNN} of noninteracting fermions
summarized in an  graph known as the Hofstadter butterfly\cite{DRH,book}. In its pure form, the Hamiltonian of such a system can be written as\cite{Wil},
\begin{equation}
H = \cos p + \lambda \cos x, 
\label{BH}
\end{equation}
where $ [x, p] = i \phi$.
In other words,  the parameter $\phi$,  the magnetic flux , measured in units of the
flux quantum $\hbar/e$,  threading every unit cell of a square lattice plays the role of the Planck constant.
This one-dimensional mathematical representation of the two-dimensional problem is a consequence of the  gauge choice  originating from the fact that  there is an inherent freedom in choosing the vector potential
that determines the magnetic field.   Within  the Landau gauge, the two dimensional wave function decouples
as $\Psi_{n,m}= e^{ i k_y m} \psi_n(k_y)$ where the Bloch vector $k_y$ appears as a phase factor $\delta$ in the above equations.

 For rational flux value $\phi=p/q$,  eigenstates of  the Harper's equation are Bloch states and the energy spectrum  consists of $q$ bands .
However,  for irrational values of $\phi$,  the system exhibits localization transition\cite{AA}:
for $ \lambda < 1$ states are Bloch states while for $\lambda > 1$ states are exponentially localized
  The critical point $\lambda=1$ has been extensively studied by various renormalization group (RG) analysis\cite{RG1, RG2}. These studies for quadratic irrational values of $\phi$ such as the golden mean describe self-similar spectral properties of the system.  
  
  Somewhat less known is the {\it strong coupling}  ($\lambda \rightarrow \infty$) fixed point of the Harper's equation, where the fractal fluctuations  about exponentially decaying wave functions exhibit self-similar characteristics with universal power law scaling\cite{KS}.
  Interestingly,  the  strong coupling limit of the Harper equation has a direct correspondence with the quasi periodic $\cal{PWS}$ chain. In $1995$ paper by Ketoja and Satija\cite{KS},
 it was pointed out that  $\lambda  > 1$,  universal properties of the Harper equation describes the onset to
 topological phase transition in the superconducting chain.  This relationship (briefly reviewed in section II below) is the key in identifying  Chern numbers of quantum Hall problem described by the Harper's equation 
 as a topological length scale associated with the Majorana wave functions in a quasi periodic $\cal{PWS}$.
 
 In this paper, we show that quasi periodicity induces oscillations in the Majorana modes  
 that are damped  as one moves in the interior of the chain.   The system was very briefly touched in our recent paper\cite{SN}
where the effect of quasi periodicity on the Majorana wave function was described as the spitting of the Majorana peaks. Here, we explore the subject in greater detail   and bring out additional richness and clarity to elucidate  the effects 
of quasi periodic disorder on the topological Majorana mode as well as the topological critical point of the superconducting chain. Our study is extended to a generalized chain which is doubly quasi periodic and exhibits a new type
of interplay between topology and quasi periodicity.

 Section III describes the Majorana  modes in the quasi periodic superconducting chain.   By studying  the edge modes that exist in a {\it hidden} dimension related to translational invariance of the quasi periodic disorder,  we show that 
 the topological length scale is the Chern number of the 2DEG problem described by the Harper's equation. The correspondence between the strong coupling limit  of Harper ( that does not alter the topological landscape
 of the system) and the critical point of the topological phase transition in $\cal{PWS}$ is the key in identifying what we refer as the {\it Chern dressing} of the Majorana wave function.
In section IV, we study an extended Harper\cite{FC,Thou} and the corresponding superconducting chain and show that the relationship between  the Majorana spreading and the
 Chern number persists below a bicritical point.  Topological and fractal characteristics of the system above this bicritical point 
are described by a  new universality class as was found to be the case also in an extended Harper  by Thouless\cite{Thou}. 
Section V examines the scaling properties  of the zero energy mode at the onset to topological phase transition in $\cal{PWS}$ beyond the bicritical point  and show that they are described by an invariant fractal set.
In appendix, we show the general mapping between the extended Harper and doubly quasi periodic $\cal{PWS}$ where diagonal hopping along two diagonals of a square lattice may not be equal.
This mapping that includes the special case of triangular lattice  reveals a kind of anomalous term in the spin Hamiltonian and may be  relevant to a earlier studies of mathematical properties
of extended Harper system\cite{GHarper, KS2}.

\begin{figure}[htbp]
\includegraphics[height=3.5in,width=3.5in]{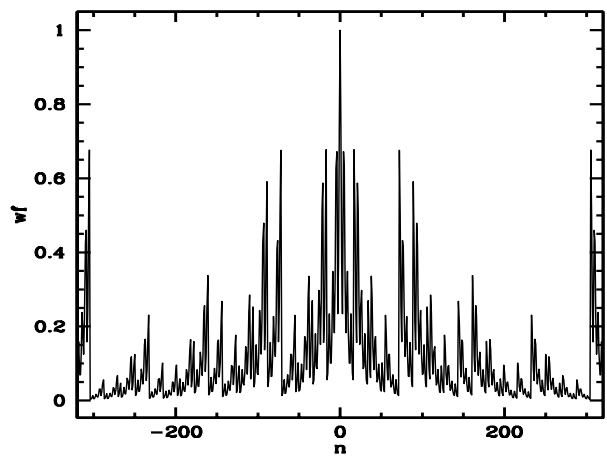}
\leavevmode \caption{ Fluctuations in $E=0$ wave function for the supercritical Harper equation that describes  the critical point of the topological phase transition.  Asymmetrical nature of various peaks  indicates
exponential decay of the Majorana modes below criticality.}\label{wfsc}
\end{figure}

\begin{figure}[htbp]
\includegraphics[height=3.0in,width=3.5in]{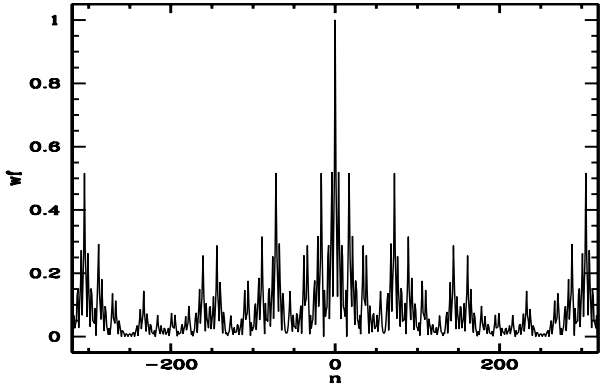}
\leavevmode \caption{ Figure shows  $E=0$ wave function at the critical point ($\lambda=1$) of the Harper's equation, illustrating a sharp contrast with the  supercritical wave function shown in Fig. (\ref{wfsc}).
At $\lambda=1$, various peaks exhibit approximately symmetrical decay  about the peak center whereas in supercritical wave functions   decay highly asymmetrically. }
\label{wfc}
\end{figure}
 
 \section{ Strong Coupling Fixed Point of Harper Equation and Topological Phase Transition in p-Wave Superconducting Chain}
 
 We briefly review the relationship between the strong coupling fixed point of  the Harper's equation and the critical point of the $\cal{PWS}$ chain describing topological phase transition.

For $\lambda > 1$, the Harper's equation for incommensurate flux exhibits localized state.  As described  in Ref.  (\cite{KS}) fractal character of  these exponentially localized states  emerge once
 we factor  out the exponentially decaying part of the wave function in Eq. (\ref{Harper}) as,
 
 \begin{equation}
  \psi_n  \equiv  e^{-n\xi} \eta_n
  \end{equation}
 
 where the localization length $\xi$ is given by, 
  
  \begin{equation}
 \xi =\frac{1}{ ln \lambda}
 \end{equation}
 
Hence $\eta_n$ describes fluctuations in the exponentially decaying localized wave function of the Harper equation. From Eq. (\ref{Harper}), it follows that the
fluctuations $\eta_n$  satisfy the following equation,
\begin{equation}
e^{-\xi} \eta_{n+1} + e^ {\xi} \eta_{n-1} + 2 \lambda \cos( 2\pi (\phi n + k_y)) \eta_n = E \eta_n
\label{fluc}
\end{equation}

Since $e^{-\xi} = \lambda$,   the limit $\lambda \rightarrow   \infty$,  leads to the following {\it strong coupling} form of the Harper equation,

\begin{equation}
 \eta_{n-1} + 2  \cos 2\pi (\phi n + k_y) \eta_n = \frac{E}{\lambda} \eta_n 
\label{etaU}
\end{equation}

We note that for $E=0$,  Eq. (\ref{etaU}) reduces to the Eq. (\ref{superc}) for E=0 at its critical point  for $\Delta =1$  as  $ \lambda_T = 1+ \Delta = 2$.
It turns out that even for $\Delta \ne 1$, the  universal aspects of the $E=0$ wave function at criticality is described by the Eq. (\ref{etaU})\cite{KS}. In other words, the strong coupling fixed point of the Harper's equation describes the critical point of the topological phase transition in $\cal{PWS}$ chains, for all values
 of the gap parameter $\Delta$.

Figure (\ref{wfsc})  shows the  self similar fractal fluctuations of the $E=0$ mode of the Harper's equation for inverse golden mean flux. We note that every sub peak is highly asymmetrical,  reminiscent of the
exponential decay of Majorana modes in the topological phase as described in next section.
This is in sharp contrast to the corresponding  self-similar  critical wave function of the Harper equation as shown in Fig. (\ref{wfc}).

\section{ Majorana Modes in  a Quasiperiodic  $\cal{PWS}$}

 \begin{figure}[htbp]
\includegraphics[height=4.0in,width=4.0in]{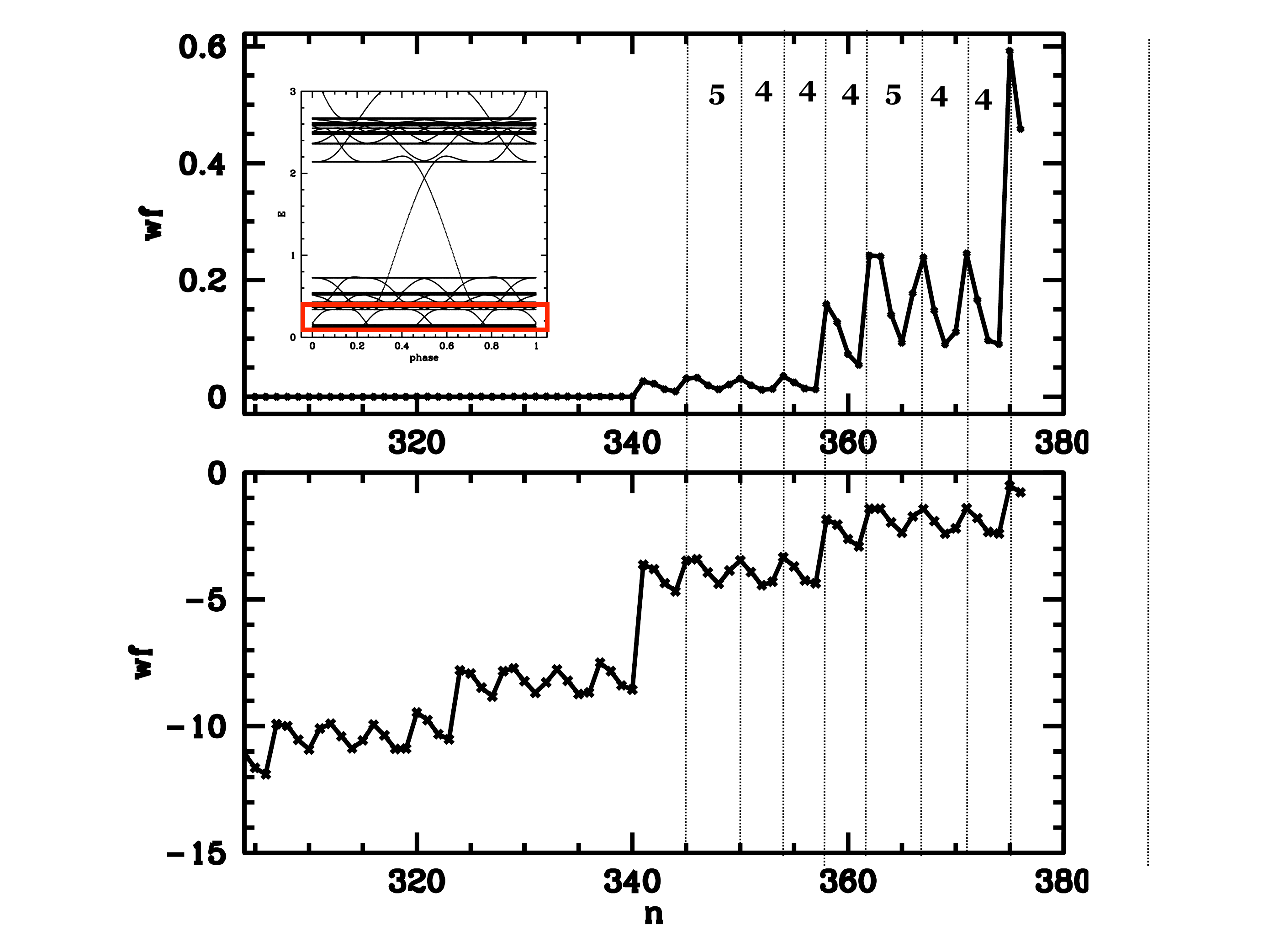}
\leavevmode \caption{(color online)  Majorana wave function at $\phi$ value  corresponding to  the inverse golden mean,
marked by  the integer $4$ in Fig. (\ref{BHI}).   It shows the ``shadowing effect" of the four edge modes both on a linear and on a log scale at $\lambda=1.75$ and $\Delta=1$.
The insert shows the energy spectrum as a function of the phase -- a hidden dimension in the quasi periodic superconducting chain where the four edge states, both at the left and the right edges are shown inside red box.}
\label{wf4}
\end{figure}

 \begin{figure}[htbp]
\includegraphics[height=3.0in,width=3.75in]{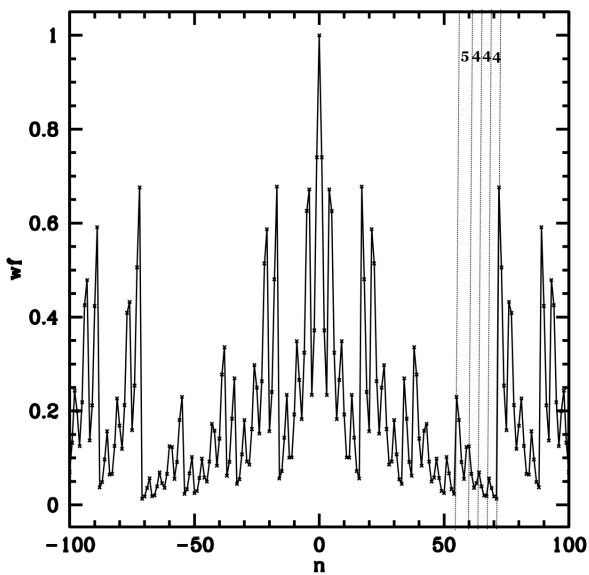}
\leavevmode \caption{(color online)  Wave function at the onset to topological transition, where the separation between the peaks exhibit same pattern as found in the topological phase as shown in Fig. (\ref{wf4}).}
\label{SCnum}
\end{figure}

We now discuss topological phase of the quasi periodic $\cal{PWS}$ characterized by Majorana modes.
 
  \begin{figure}[htbp]
\includegraphics[height=3.5in,width=3.8in]{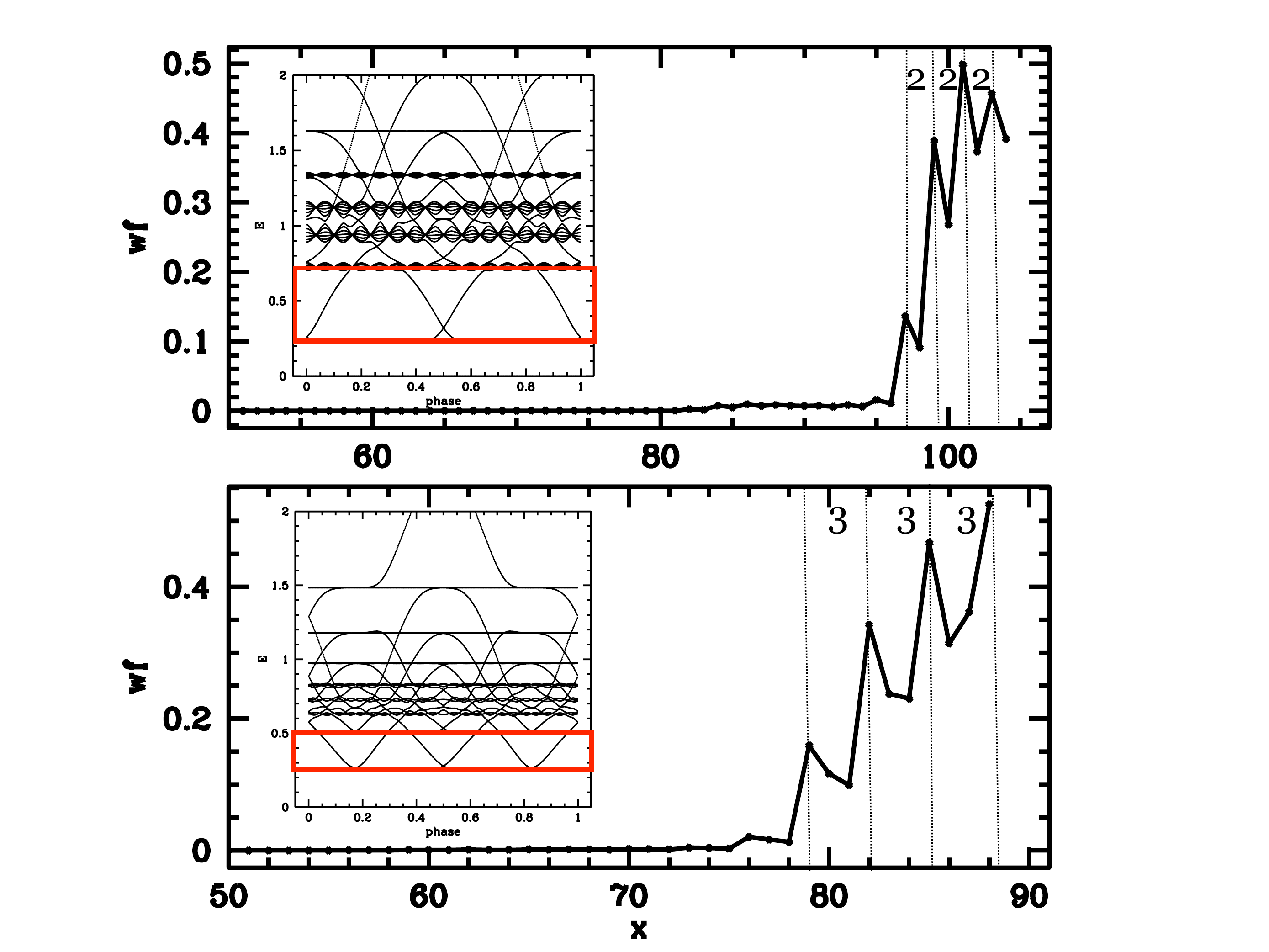}
\leavevmode \caption{(color online)  Analogous to Fig. (\ref{wf4}),   this graph shows the Majorana wave function exhibiting  oscillatory behavior with period $2$ ( upper)  and $3$ ( lower)   at $\lambda=1.75$ and $g=1$ at two other values
of $\phi$ that are marked in Fig. (\ref{BHI})
 by the integers $2$ and $3$ representing the Chern numbers  at the  major gap near $E=0$.
The insets show
the corresponding edge modes.}
\label{wf2and3}
\end{figure}

\begin{figure}[htbp]
\includegraphics[height=6 in,width=3.5 in]{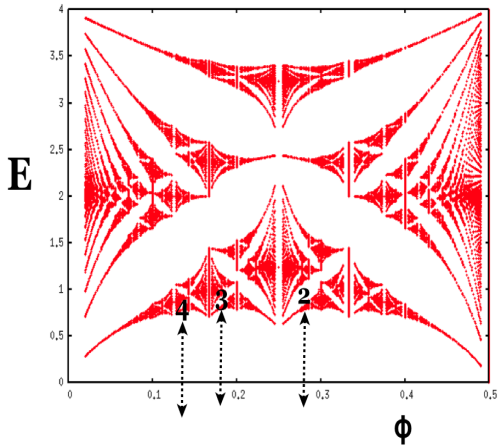}
\leavevmode \caption{(color online)   Spectral graph resembling Hofstadter butterfly for the  superconducting chain with $\Delta=1$ at subcritical value of $\lambda =1.5$.
The three arrows  show the incommensurability parameter $\phi$ corresponding to the figures   (\ref{wf4}) and (\ref{wf2and3}) ( top and bottom),  labeled with integers $4$, $2$ and $3$.
 Fig. (\ref{Edelta}) below justifies the integer labeling of the gaps
in the superconducting butterfly.}
\label{BHI}
\end{figure}

\begin{figure}[htbp]
\includegraphics[height=3.1in,width=3.4in]{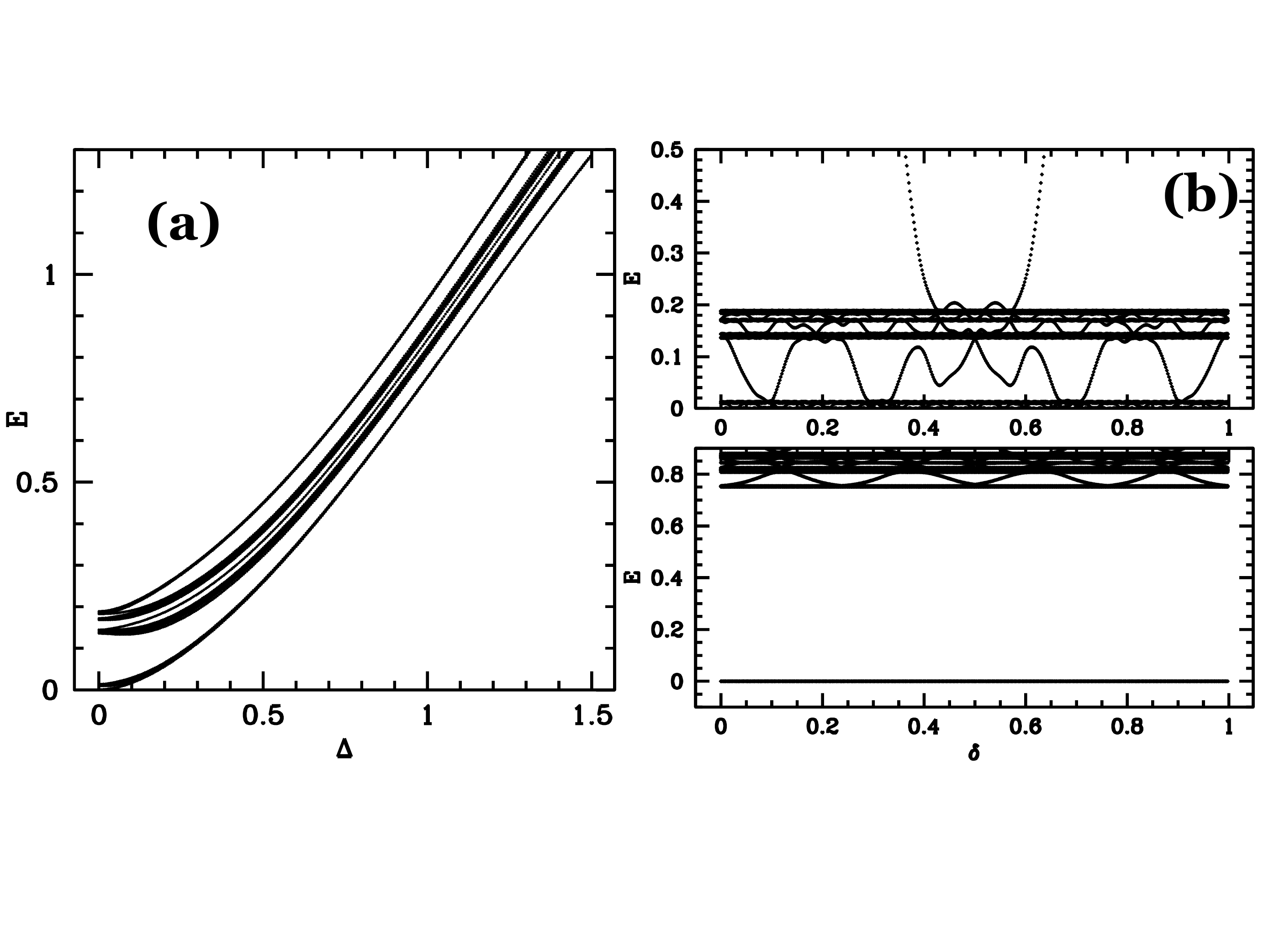}
\leavevmode \caption{(color online)  (a) Energy spectrum near $E=0$ as a function of the gap parameter $\Delta$ for a fixed value of $\phi$. Figure illustrates the fact that  by varying $\Delta$,
the topological character of the spectrum near
$E=0$ remains unchanged  as there are no band crossings. This justifies labeling the superconducting butterfly with the Chern numbers of the Hofstadter butterfly.
 (b) Energy spectrum  showing edge modes for Harper (top) and superconducting topological phase ( bottom) near $E=0$. }
\label{Edelta}
\end{figure}

Analogous to a periodic chain, the sinusoidally modulated quasi periodic $\cal{PWS}$  with $\mu_n$ given by Eq. (\ref{qumu}) supports  Majorana excitations
 for $\lambda < \lambda_T = 1+\Delta$.
 Fig. (\ref{wf4}) shows  the Majorana wave function,  for $\phi$ equals to the inverse golden mean that we denote as $\phi_g$.  In sharp contrast to a periodic chain,
 the edge states penetrate the bulk
exhibiting damped oscillations characterized by a spatial periodicity.  The inserts in these graphs are the corresponding energy spectrum   showing edge modes near $E=0$.
It should be noted that these spectral insets show energy as a function of the phase parameter $\delta$ and in that sense the edge states exist in a fictitious dimension. The key point to be noted here is that the {\it number of the edge modes near $E=0$
coincide with the spatial periodicity of the oscillating Majorana wave function}. In other words, the spatial length scale characterizing recurrence of damped oscillations in the Majorana mode has  a topological origin. 

 The damped Majorana peaks are spaced with a spatial pattern of $4,4,4, 5$ that continues throughout the chain as highlighted in the log scale graph. The emergence of spacing  $5$ after three successive spacings of $4$ is an example
 of competing length scales as the topological length $4$ tries to ``adapt" to quasi periodic pattern which for the golden mean ( $\phi = \phi_g$)  case leads to peaks separated by rational approximates of $\phi_g^3$\cite{RG1}.  This type of competition between topology and quasi periodic was  discussed in  in our earlier paper\cite{SN}. We also note that the characteristic pattern of the Majorana wave function is identical to the pattern seen at the topological transition point
 as shown in Fig. (\ref{SCnum}).
 
Fig (\ref{wf2and3})  shows the  analogous spatial profiles of the Majorana for two other values of the commensurability parameter $\phi$, confirming the relationship between the spatial period associated with the Majorana and the  number of edge modes in the spectral graph.

 We next argue that this topological length scale encoded in the oscillatory profile of the Majorana mode is related to the Chern numbers of the  two dimensional electron gas (2DEG) problem described the Harper's equation\cite{QHE}.
 The Harper spectrum evolves smoothly into the quasiparticle excitation spectrum of the superconducting chain, particularly near $E=0$ as there are no band crossings as shown in the Fig. \ref{Edelta}). This allows us to assign integers to the gaps of the superconducting butterfly spectrum. Figure  \ref{Edelta} also shows  the correspondence between edge modes\cite{Hat} in Harper and $\cal{PWS}$  for $\phi$ close to the inverse golden-mean where the Chern $4$ state is the dominant state near $E=0$ in the Harper equation and we see four edge modes in Harper as well as in the superconducting chain as illustrated in the figure.
 Unlike Harper equation where the number of edge modes represent the topological quantum number of Hall conductivity,
 the physical significance of the topological integer in the Ising chain or $\cal{PWS}$ remains elusive, specially in view of the fact that edge states  exists in a fictitious dimension.

  Remarkable fact that the Majorana modes are shadowed by the edge modes existing  in a nearby gap in a hidden dimension encoded in the phase factor $\delta$ is consistent with the reasoning that emerged from
  recent studies\cite{QCti}  arguing the fact that  quasicrystals  are topological states of matter.
  This  topological character was unveiled  by edge modes existing in a fictitious dimension -- the phase factor,  related to the  
translational invariance that shifts the origin of quasiperiodicity and manifests as an
additional degree of freedom . This {\it hidden} degree of freedom was used to relate the quasi periodic systems
 to higher dimensional periodic systems\cite{QCti} and associate topological feature to quasi periodic systems, assigning  topological insulator\cite{TI} status to quasi periodic systems.
An explicit demonstration of edge transport in the quasicrystals  mediated by the edge modes, was demonstrated by pumping light across photonic quasicrystals\cite{QCT2}.

\section{Extended Harper and doubly Quasiperiodic Superconducting Chain}

We next consider a generalized Harper\cite{FC,Thou}, commonly referred as the {\it extended Harper} described by the Hamiltonian,
\begin{equation}
H^{ext} = \cos x +  \lambda \cos p + \alpha \cos ( p+x) +  \beta \cos(p-x), 
\label{Hext}
\end{equation}

   \begin{figure}[htbp]
\includegraphics[height=3in,width=3.5in]{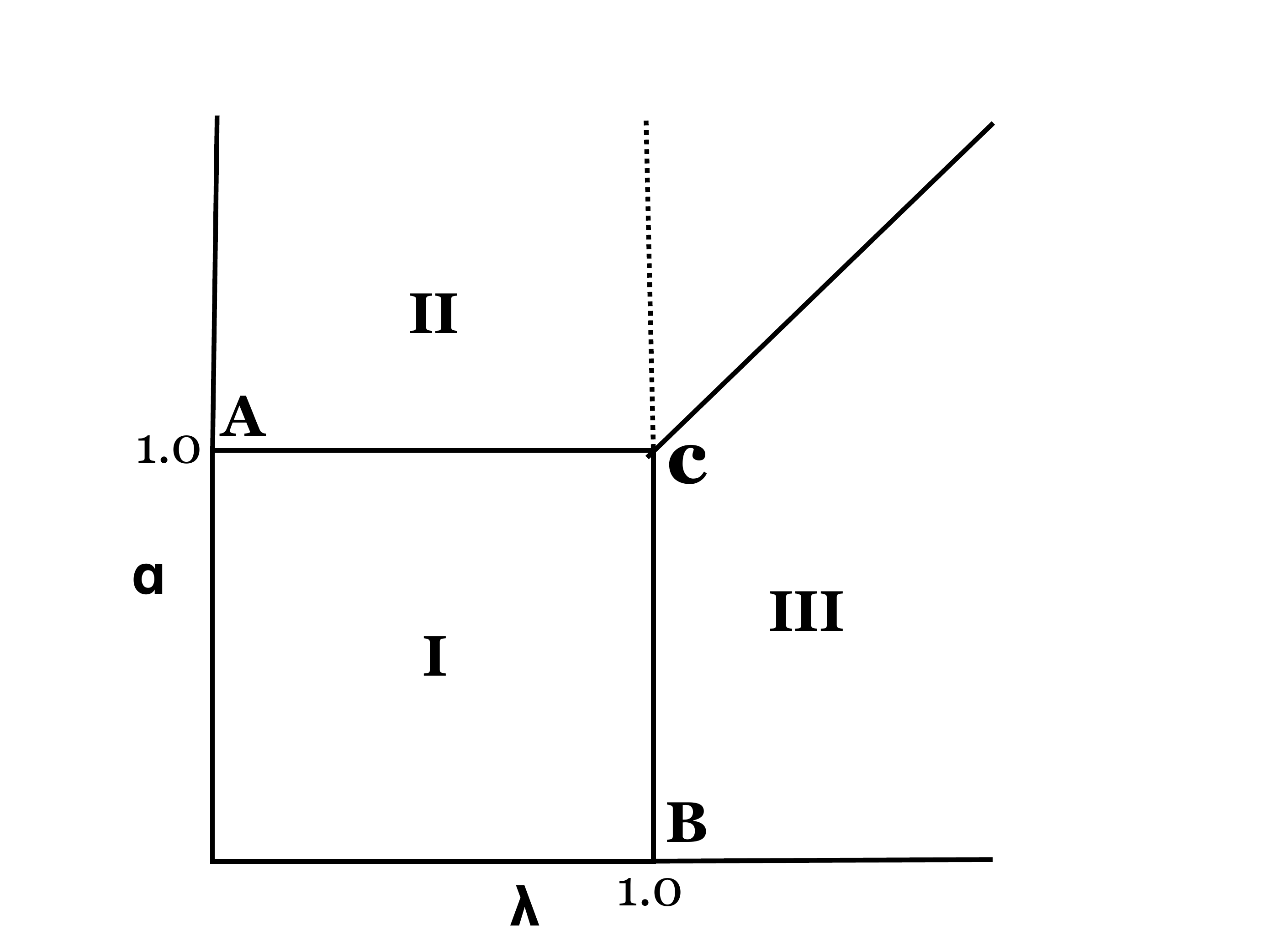}
\leavevmode \caption{(color online)  Schematic phase diagram as a function of $\lambda$ and $\alpha=\beta$ for extended Harper exhibiting three distinct types of phases labeled as $I$, $II$ and $III$ 
for incommensurate $\phi$.   Regime $I$ shows the extended phase while the regime II is localized phase. In the regime III, states are critical where the wave function and spectra are multi fractal.
For $\lambda \le \alpha$ the model belongs to Harper university class.  The point $C$ is the bicritical point and for $\alpha > \lambda$, spectral characteristics are different from those of Harper model, both at the critical and strong coupling points. As discussed below, for parameters above the bicritical point,  the correspondence between the  number of edge modes and the recurrence period of the damped Majorana peaks is lost.}
 \label{ThouPD}
\end{figure}

where $[x, p] = i \phi$.  The parameters $\alpha$ and $\beta$ describe the strengths of the hoping along the two diagonals of the rectangular lattice, in units of nearest-neighbor hopping. For $\beta=0$, it describes the system in a triangular lattice.  
The eigenvalue equation for the extended Harper can be written as,

\begin{widetext}
\begin{eqnarray*}
[1+ 2 \alpha   \cos (2 \pi   \phi (n-1/2)-  k_y)] \psi_{n-1}  +2 \lambda \cos ( 2 \pi  \phi n + k_y) \psi_n
+[1+ 2 \alpha  \cos (2 \pi  \phi (n+1/2)- k_y)] \psi_{n+1}
 = E \psi_n
\end{eqnarray*}
\end{widetext}

 Fig. (\ref{ThouPD}) shows the schematic phase diagram  of the extended model\cite{Thou} for irrational flux  values  with $\alpha=\beta$.
For $\lambda=1$ and $\alpha=\beta$, system has square symmetry  but there are two quite different regimes with this symmetry, according to whether  $\lambda $ or $\alpha$ dominates
exhibiting an interesting bicritical point   at $\lambda=\alpha=1 $ that separates the two different regimes.   Below we discuss only this symmetric case with $\alpha = \beta$. 

For $\alpha  < 1 $, the system belongs to the university class of the Harper equation. However, above the bicritical point, we see critical behavior characterized by a Cantor set spectrum existing
in a parameter space of finite measure. Labeled as the regime II in the Fig. (\ref{ThouPD}), this regime of ``spectral collapse"  has been the subject of detailed mathematical analysis.\cite{GHarper}

In close analogy with our analysis of a relationship between the Harper equation and the superconducting chain as described earlier,  we now seek a   generalized superconducting  or spin chain that may
relate to the extended Harper model.  
Firstly, we note that using Jordan Wigner transformation, the  extended Harper model equation can be shown to map to a one-dimensional spin chain, given by the following  Hamiltonian,

\begin{eqnarray*}
H^{ext}_0 =\sum_{ n = -\infty} J_n ( \sigma^x_n \sigma^x_{n+1}
  + \sigma^y_{n} \sigma^y_{n+1})
  + 2 \lambda \cos (2 \pi \phi n+ k_y)  \sigma^z_n 
\end{eqnarray*}

where $J(n)= (1+ 2\alpha \cos (2 \pi \phi n+\delta))$.
This shows that  the extended Harper system  describes in fermionic representation,  a spin chain where both the exchange  spin interaction as well as the magnetic field are quasi periodic.
Following the analogy between the $\cal{PWS}$ problem and the Harper equation as discussed earlier, we now introduce  the parameter $\Delta$ in the extended spin model 
that describes spin space anisotropy.

\begin{widetext}
\begin{eqnarray}
H^{ext}_{spin} =  \sum_{ n = -\infty}  [ (J_n - \Delta) \sigma^x_n \sigma^x_{n+1}
 +  (J_n+\Delta) \sigma^y_{n} \sigma^y_{n+1})+
 2 \lambda \cos (2 \pi \phi n+k_y)  \,\  \sigma^z_n ]
\end{eqnarray}
\end{widetext}

The  resulting tight binding model, corresponding to this spin chain is,  a coupled set of equations involving two component wave functions $(\psi^a, \psi^b)$. 

\begin{eqnarray*}
(J_{n-1} +  \Delta) \psi^a_{n-1} + (J_n+\Delta)  \psi^a_{n-1} + 2  \mu_n \psi^a_n  & = & E \psi^b_n\\
(J_{n-1} - \Delta) \psi^b_{n-1} +  (J_{n} +  \Delta) \psi^b_{n-1} + 2  \mu_n \psi^b_n  & = & E \psi^a_n
\label{Gsc}
\end{eqnarray*}

These coupled set of equations   describe a superconducting chain where both the nearest-neighbor fermion hoping as well as the chemical potential are quasi periodic.

Our detailed study of this model shows a superconducting chain exhibiting  topological phase supporting Majorana modes that delocalize at a critical point $\lambda_T^{ext}$ of the transition that depends on both
$\Delta$ and $\alpha$.
In the topological phase, these damped modes exhibit a characteristic topological length scale only below the bicritical point, as shown in the figure.  Above this bicritical point,  the Majorana oscillations exhibit a random pattern, that is,  various
 sub peaks  are spaced  somewhat randomly. This feature persists also characterizes  the critical point of the topological phase transition above the bicritical point.

\section{ Universal ``Bow-Tie" Invariant Set}

We now show a kind of {\it hidden order} in the random oscillations in $E=0$ wave function in the extended $\cal{PWS}$ chain  above the bicritical point at the onset to topological phase transition by studying the strong coupling limit of the
extended Harper.

\begin{figure}[htbp]
\includegraphics[height=3.8in,width=3.6in]{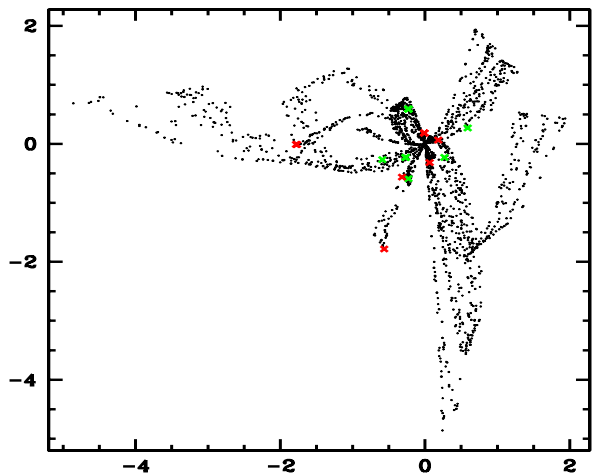}
\leavevmode \caption{(color online)   For $\phi$ equals to the inverse golden mean,  each point in the  figure is of  amplitude of  critical the wave function  for the extended model
 at two consecutive Fibonacci sites.
With $ \alpha > 1$  trajectories at various  Fibonacci sites converge on an invariant set ( shown in black).
We note that points corresponding to low Fibonacci numbers represents ``transients" are not plotted on the graph.
 The six cycles in red and green correspond to $\alpha < 1$  ( strong coupling Harper universality) and $\alpha=1$ ( bicritical point of strong coupling Harper). }
 \label{SA}
\end{figure}

Following Ref.\cite{KS}, we describe the universal scaling properties of the zero energy wave function for the generalized superconducting chain at the onset to topological transition  
 by monitoring the amplitude of the wave function at  the Fibonacci sites where $\phi$ is chosen to be the inverse golden mean. System exhibits a periodic cycle of period six\cite{KS}
 as long as the parameter space is below or at the bicritical point as shown with two distinct cycles (in red and green) in Fig. (\ref{SA}) .
 
  However, above the bicritical point,   the amplitudes  of the wave function at Fibonacci sites vary randomly.
 Interestingly, the random set of amplitudes converge on an invariant set as we vary $\alpha > 1$. Reminiscent of the ``orchid flower" ( that has also been analyzed  mathematically\cite{Orchid}  ),
 for the band end energy\cite{KS},
 the invariant set for $E=0$ state has a shape of a bow-tie. This shows an intriguing example of order and complexity at the topological critical point when subjected to quasi periodic disorder in both the hopping as well as the chemical potential.
 
\section{ Summary}

Damped oscillations of the Majorana modes in a quasi periodic $\cal{PWS}$ chains where the period of oscillations has topological root describes a novel interplay between two distinct types of topologies.
 Interestingly, this Chern {\it dressing} of the wave function  persists even at the onset to the topological phase transition. 
Such  ``trail marks"  left behind by the Majorana as it disappears at the critical point may facilitate in the detection of Majorana modes in laboratories\cite{MFexpt}. 
Our study of a generalized model where scaling associated with the multi fractal wave functions at the topological phase transition are described by a strange attractor raises interesting mathematical question about 
the role of topological states in determining such stable strange sets.
\appendix

\newpage

\section{ Spin Mapping for Extended Harper with $\alpha \ne \beta$}

We now describe spin-Fermion mapping for the extended Harper for $\alpha \ne \beta$. The generalized tight binding model can be written as,

\begin{widetext}
\begin{eqnarray*}
[1+\alpha e^{- 2 \pi  i \phi (n-1/2)- i k_y}+\beta e^{- 2 \pi  i\phi (n-1/2)+i k_y}] \psi_{n-1}  +2 t_b \cos ( 2 \pi  i\phi n + k_y) \psi_n\\
+[1+\alpha e^{- 2 \pi  i\phi (n+1/2)- i k_y}+\beta e^{- 2 \pi  i\phi (n+1/2)+i k_y}] \psi_{n+1}
 = E \psi_n
\end{eqnarray*}
\end{widetext}

Just like Harper model, we now explore the corresponding spin and superconducting chains  hidden in this fermionic Hamiltonian.

Using Jordan Wigner transformation, the equation can be shown to map to the  one-dimensional spin chain, given by the following  Hamiltonian,

\begin{widetext}
\begin{eqnarray*}
H_s &= &\sum_{ n = -\infty}  \frac{1}{2} [ 1 + ( \alpha+\beta) \cos (2 \pi \phi n+k_y) ] ( \sigma^x_n \sigma^x_{n+1}
 + \sigma^y_{n} \sigma^y_{n+1})\\
  &+& \frac{1}{2} ( \alpha-\beta)  \sin (2 \pi \phi n+k_y) \,\
( \sigma^x_n \sigma^y_{n+1} - \sigma^y_{n} \sigma^x_{n+1}) + 2 \lambda \cos (2 \pi \phi n+k_y)  \,\ \sigma^z_n
\end{eqnarray*}
\end{widetext}

Note that for $\alpha \ne \beta$,  the spectral collapse disappears\cite{KS2,GHarper}. This mapping may be relevant in recent  mathematical studies (\ref{Harper}) exploring the the absence of Cantor sets of zero measure existing
in extended Harper model.

\end{document}